\documentclass[10pt,journal,compsoc]{IEEEtran}

\ifCLASSOPTIONcompsoc
  \usepackage[nocompress]{cite}
\else
  \usepackage{cite}
\fi

\ifCLASSINFOpdf

\else
\fi

\usepackage[T1]{fontenc}
\usepackage{txfonts}
\usepackage[pdftex]{hyperref}
\usepackage{color}
\usepackage{siunitx}
\usepackage{booktabs,arydshln}
\usepackage{textcomp}
\usepackage{multirow}
\usepackage[caption=false]{subfig}
\usepackage{microtype} 
\usepackage{ccicons}  
\usepackage{mathptmx}
\usepackage{graphicx}
\usepackage{times}
\usepackage{enumitem}
\usepackage{tabu}
\usepackage{cite}

\makeatletter
\def\adl@drawiv#1#2#3{%
        \hskip.5\tabcolsep
        \xleaders#3{#2.5\@tempdimb #1{1}#2.5\@tempdimb}%
                #2\z@ plus1fil minus1fil\relax
        \hskip.5\tabcolsep}
\newcommand{\cdashlinelr}[1]{%
  \noalign{\vskip\aboverulesep
           \global\let\@dashdrawstore\adl@draw
           \global\let\adl@draw\adl@drawiv}
  \cdashline{#1}
  \noalign{\global\let\adl@draw\@dashdrawstore
           \vskip\belowrulesep}}
\makeatother

\begin{document}

\title{Task-Based Effectiveness of Basic Visualizations}

\author{Bahador~Saket,
        Alex~Endert, and \c{C}a\u{g}atay Demiralp
\IEEEcompsocitemizethanks{\IEEEcompsocthanksitem Bahador Saket and Alex Endert are
with Georgia Tech.\protect\\
E-mail: \{saket, endert\}@gatech.edu.

\IEEEcompsocthanksitem \c{C}a\u{g}atay Demiralp is with IBM Research. \protect\\
E-mail:cagatay.demiralp@us.ibm.com \textbf{}
}
\thanks{Manuscript received April 19, 2005; revised August 26, 2015.}}

\markboth{Journal of \LaTeX\ Class Files,~Vol.~14, No.~8, August~2015}%
{Saket \MakeLowercase{\textit{et al.}}: Task-Based Effectiveness of Basic Visualizations}

\IEEEtitleabstractindextext{%
\begin{abstract}
Visualizations of tabular data are widely used; understanding their effectiveness in different task and data contexts is fundamental to scaling their impact. However, little is known about how basic tabular data visualizations perform across varying data analysis tasks. In this paper, we report results from a crowdsourced experiment to evaluate the effectiveness of five small scale (5-34 data points) two-dimensional visualization types---Table, Line Chart, Bar Chart, Scatterplot, and Pie Chart---across ten common data analysis tasks using two datasets. We find the effectiveness of these visualization types significantly varies across task, suggesting that visualization design would benefit from considering context-dependent effectiveness. Based on our findings, we derive recommendations on which visualizations to choose based on different tasks. We finally train a decision tree on the data we collected to drive a recommender, showcasing how to effectively engineer experimental user data into practical visualization systems. 

%
%
\end{abstract}

}

\maketitle

\IEEEdisplaynontitleabstractindextext

%
\IEEEpeerreviewmaketitle

\ifCLASSOPTIONcompsoc

\IEEEraisesectionheading{\section{Introduction}\label{sec:introduction}}
\else
\section{Introduction}
\label{sec:introduction}
\fi

\IEEEPARstart{V}{isualizations} aim to enhance understanding of underlying data by leveraging visual perception, evolved for fast pattern detection and recognition. Understanding the effectiveness of a given visualization in achieving this goal is a fundamental pursuit in visualization research and has important implications in practice.
  
  

A large body of prior research evaluated the general effectiveness of different visualization types~\cite{eells1926relative,croxton1927bar,spence1991displaying, garcia2010profits,harrison2014ranking,correll2014error,kay2016ish}. Guidelines and insights derived from these earlier studies have significant influence on data visualization today. However, these studies were conducted under conditions that were inconsistent across studies, with varying sample sizes, a limited number of tasks, and using different datasets. Research indicates, however, the effectiveness of a visualization depends on several factors including task at the hand~\cite{amar2005low}, and data attributes and datasets visualized~\cite{santos2008evaluating}. For example, while one chart might be suitable for answering a specific type of question (e.g., to check whether there is a correlation between two data attributes), it might not be appropriate for other types (e.g., to find a data point with the highest value). Yet, we know little about how some of the basic visualizations perform across different visual analysis tasks. 

In this paper, we conducted a crowdsourced study to evaluate the effectiveness of five small scale (5-34 data points) two-dimensional visualization types (Table, Line Chart, Bar Chart, Scatterplot, and Pie Chart) across 10 different visual analysis tasks~\cite{amar2005low} and from two different datasets (Cars and Movies). Our results indicate that the effectiveness of these visualization types often significantly varies across tasks. For example, while pie charts are one of the most effective visualizations for finding the extremum value, they are less effective for finding correlation between two data attributes. We also asked participants to rank five different visualization types in the order of their preference for performing each task. We found a positive correlation between accuracy and user preference, indicating people have a preference for visualizations that allow them to accurately complete a task. 

There is a renewed interest (e.g.,~\cite{vartak2014seedb, bouali2015vizassist, Kandel:2012, saket2017visualization, 2015-voyager, Mackinlay:2007}) in visualization recommendation systems that aims to shift some of the burden of visualization design and exploration decisions from users to algorithms. Our results can be used to improve visualization recommendation systems moving forward. In particular, 
our findings from the current study inform the ongoing design and development of Foresight~\cite{demiralp2017foresight} at IBM. 
We envision creating a recommendation engine that suggests visualizations based on user-specified tasks. To this end, we develop \textbf{Kopol}\footnote{\url{https://kopoljs.github.io/}}, a prototype visualization recommender. A decision tree model is trained on the user data and then used by Kopol to provide ranked recommendations for a given task and data type. This model takes into account performance time, accuracy, and user preference. One relevant application area of such a recommendation engine can be natural 
language interfaces for data visualization (e.g.,~\cite{articulate, eurovisshort.20171133}). In such interfaces people tend to specify 
tasks as a part of their queries (e.g., ``Is there a correlation between price and width of cars in this dataset?''). 
Such an engine can be used to suggest more effective visualizations given the task context.
 
  
\section{Related work}

Data representation is a main component of information visualizations.
The fundamental focus of data representation is mapping from data values to
graphical representations~\cite{cleveland1985graphical}. Visualization designers use elementary graphical units
called visual encodings to map data to graphical representation. Through human-subject experiments, researchers have investigated the effects of visual encodings on the ability to read and make judgments about data represented in visualizations (e.g.,~\cite{bertin:book83, skau2016arcs, simkin1987information,spence1991displaying, saket2017evaluating}). 

Although prior research has proposed models of visualization comprehension~\cite{kosslyn1989understanding, pinker1990theory, simkin1987information}, little is known about how visual encodings or design parameters interact with each other or different data and task contexts in
forming the overall performance of a given visualization. A large body of earlier
work (e.g.,~\cite{eells1926relative,croxton1927bar,spence1991displaying, garcia2010profits,skau2016arcs, harrison2014ranking,correll2014error,kay2016ish, dambacher2016graphs, Dimara}) has also studied the effectiveness of visualization types with common design configurations for a selected number of tasks. 


Eells~\cite{eells1926relative} investigated effectiveness, of proportional
comparison (percentage estimation) task in divided (stacked) bar charts and
pie charts. Eells asked participants to estimate the proportions in pie charts
and bar charts. He found pie charts to be as fast as and more accurate than bar
charts for proportional comparison tasks. He also found that as the number of
components increases, divided bar charts become less accurate but pie charts become
more (maximum five components were considered). In a follow up study with
a different setting, Croxton and Stryker~\cite{croxton1927bar}
also tested the effectiveness of divided bar charts and pie charts using a proportional
comparison task. They also found pie charts to be more accurate than divided bar charts
in most cases, but contrary to Eells' study, not all.

Spence et al.~\cite{spence1991displaying} studied the effectiveness of bar
charts, tables and pie charts. They found that when participants were asked to
compare combinations of proportions, the pie charts outperformed bar charts.
Their results also show that for tasks where participants were asked to retrieve
the exact value of proportions, tables outperform pie charts and bar charts.
In another study comparing the effectiveness of bar charts and line charts,
Zacks and Tversky~\cite{zacks1999bars} indicated that when participants were
shown these two types of visualizations and asked to describe the data, they
constantly used bar charts to reference the compared values (e.g., A is
10\% greater than B). Whereas with line charts, participants described trends.

Study by Siegrist~\cite{siegrist1996use} was one of the first studies that
compared 2D with 3D visualizations. Siegrist found that there is not a
significant between 2D and 3D bar charts in terms of accuracy. However,
participants using 3D bar charts take slightly longer to perform tasks. In
addition, Siegrist found that accuracy of perceiving 3D pie charts is
significantly lower than 2D ones, probably because some of the slices in the
3D pie charts are more obscured. Harrison et al.~\cite{harrison2014ranking} measured the effectiveness of different visualizations for explaining correlation, finding that parallel coordinates and scatterplots are best at showing correlation. They also found
that stacked bar charts outperform stacked area and stacked line. In a follow
up study, Kay and Heer reanalyzed~\cite{2016-beyond-webers-law} the data
collected by Harrison et al.~\cite{harrison2014ranking}. The top ranking visualization remained the same.



\begin{figure*}
    \centering
       \includegraphics[width=0.19\textwidth]{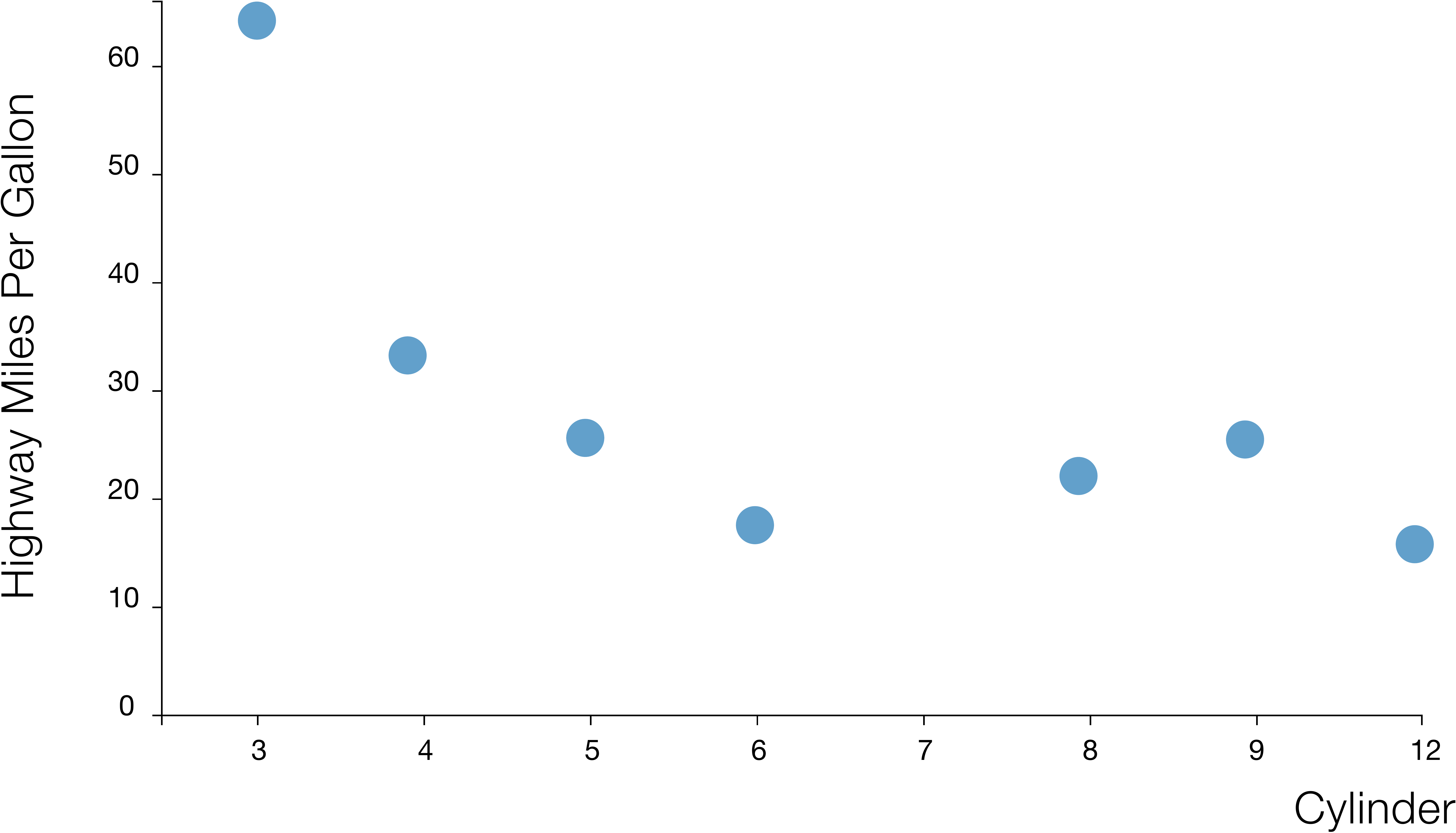}\hspace{1em}
       \includegraphics[width=0.19\textwidth]{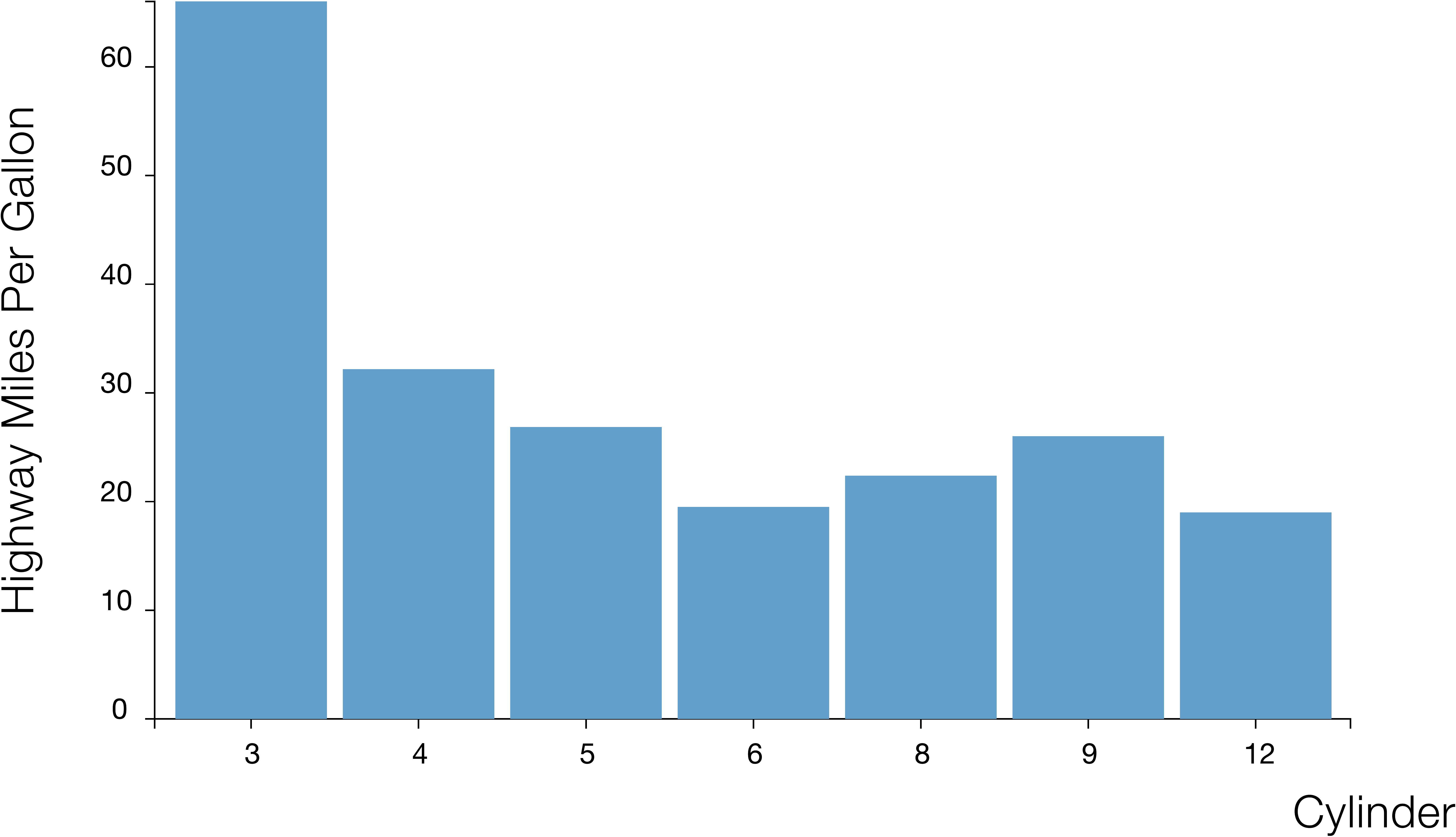}\hspace{1em}
       \includegraphics[width=0.19\textwidth]{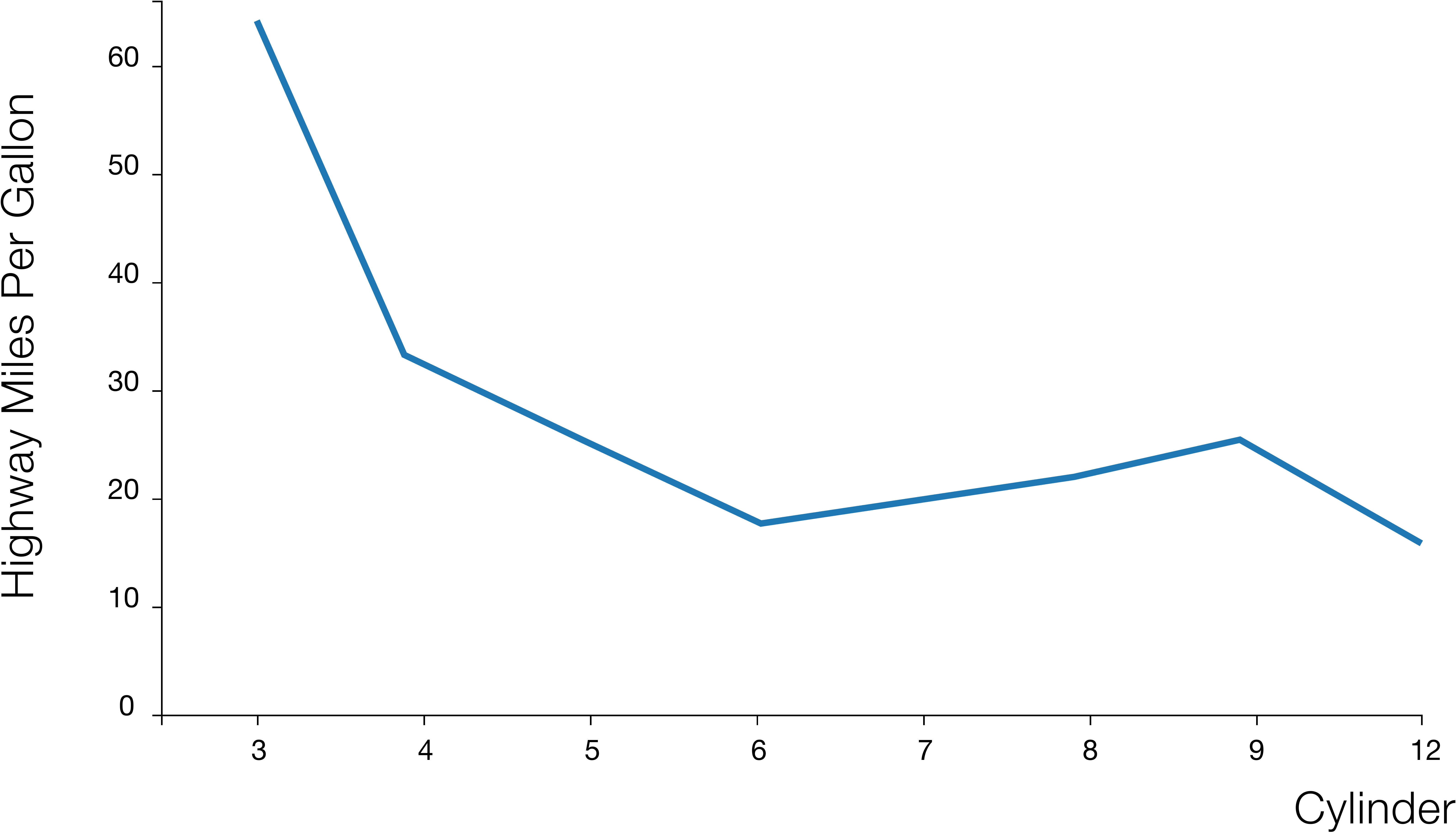}\hspace{1em}
       \includegraphics[width=0.18\textwidth]{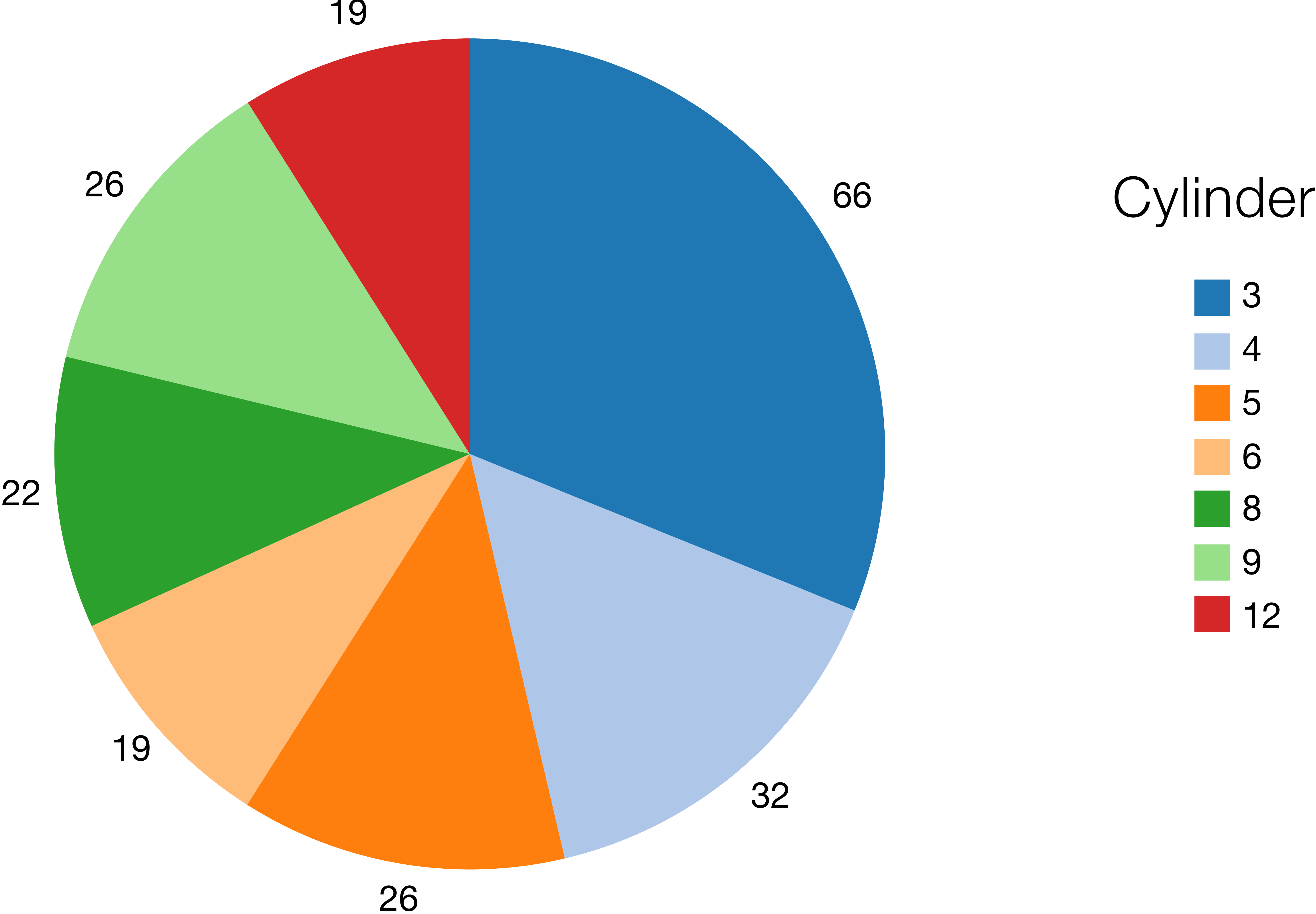}\hspace{1em}
        \includegraphics[width=0.13\textwidth]{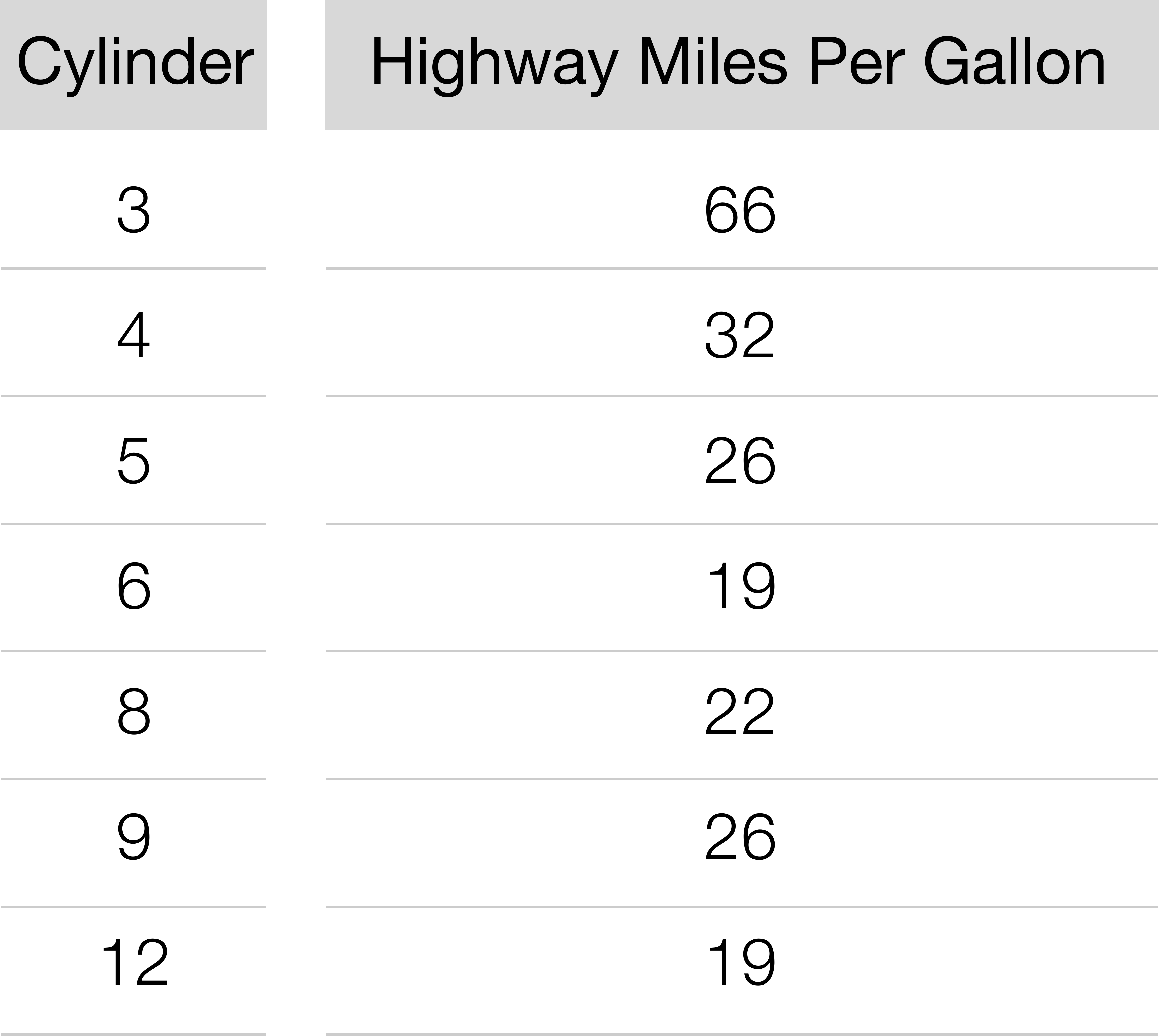}
    \caption{Five visualization types used in this study. In this figure, each visualization shows the average highway miles per gallon (a numerical data attribute) for cars with different number of cylinders (an ordinal data attribute). }~\label{fig:vis}
 
\end{figure*}

While these independent studies provide helpful generic guidelines, they were
conducted under different conditions, varying sample sizes, datasets, and for a disperse set of tasks. In fact, several of these studies used manually created visualizations in their experiments without using actual datasets~\cite{eells1926relative, croxton1927bar, spence1991displaying, zacks1999bars} or created visualizations using artificial datasets~\cite{harrison2014ranking}. Also, these earlier studies have conducted experiments typically using atomic generic tasks such as comparison
of data values (e.g.,~\cite{zacks1999bars, croxton1927bar}) or estimation of proportions (e.g.,~\cite{eells1926relative,spence1991displaying,skau2016arcs}). However, many visual analysis
tasks (e.g., filtering, finding clusters) require integration of results from
multiple atomic tasks, limiting the applicability of earlier findings~\cite{amar2005low, Amar:2004}. Inconsistency in experimental settings and limited atomic tasks used in previous work encourages studying the effectiveness of visualization types for larger spectrum of tasks in a more consistence setting.

\section{Study Design}

When deciding which visualization types to include in our experiment, we balanced the familiarity of the visualizations considered
with the comprehensiveness of the experiment. On the one hand, we would like to
have more generalizable results, which suggested considering a broad set of
visualization techniques in our experiment. At the same time, we would like
our study to have the members of the general public as our participants:
this would suggest to include a set of visualization techniques which are
understandable by all participants. Building on previous work~\cite{LeeVlat} and investigations on 
visualization techniques supported by different visualization tools (e.g., Microsoft Excel, Tableau, Spotfire, QlikView, Adobe Analytics, IBM Watson Analytics), we
decided to include five well-recognized visualization techniques in our
study. In this study, we include Bar Chart, Line Chart, Scatterplot, 
Table, and Pie Chart (see Figure~\ref{fig:vis}).

\subsection{Datasets}

\textbf{Selecting Datasets:}
To create visualizations for our experiment, we selected datasets where the  participants were unfamiliar with the content, but familiar with the meaning of the data attributes used in the dataset. This is particularly important since we did not want user performance to be affected by how familiary participants are with the meaning of the data attributes.

We first selected five different datasets: Cereals~\cite{UCIDataset}, Cars~\cite{henderson1981building}, Movies~\cite{TableauData}, Summer Olympics Medalists~\cite{TableauData}, and University Professors~\cite{UCIDataset}. We then printed a part of each dataset on paper and showed them to six
pilot participants (4 male, 2 female). We asked participants \textit{``Please look at data attributes used in each of these datasets. Which datasets do you feel contain data attributes that you are more familiar with?''} Cars and Movies datasets were the ones that five out of the six participants selected. The Cars dataset~\cite{henderson1981building} provides details for 407 new cars and the Movies dataset~\cite{TableauData} provides details for 335 movies released from 2007 to 2012.

\textbf{Data Attribute Types:}
Both datasets include data attributes of Nominal, Ordinal, and Numerical types. We define Nominal data attribute type as categorically discrete data such as types of cars (e.g., Sedan, SUV, Wagon). Ordinal is defined as quantities within a specific range that have a natural ordering such as rating of movies (the number of unique data values ranged from 6 to 12). We define Numerical as continuous numerical data such as Profit values of movies. We generated visualizations using pairwise combinations of all three types of data attributes available in our datasets (e.g., Nominal * Numerical or Ordinal * Numerical).

\textbf{Data Sampling:}
During our informal pilot study, we generated visualizations representing different number of data points ranging from 50 to 300, in increments of 50 data points. In these visualization each visual mark (e.g., a circle or a bar) represented a data point. We noticed our pilot participants faced two challenges using static visualizations containing more than 50 visual marks. First, participants had difficulties performing some of the tasks (e.g., compute derived value and characterized distribution) using static visualizations (error rate increased and in some cases participants gave up). In addition, in some cases participants had to spend more than two minutes performing the tasks. Due to practical limitations of conducting the study (e.g., length and complexity of the experiment) with a high number of visual marks, we decided to not show more than 50 visual marks at the time. We had two options for not showing all the data points in our datasets.

First, we could pick a subset of data points and create visualizations using only that subset. In that case, each visual mark would represent a data point in our dataset. We could then create a bar chart showing manufacturers on the x-axis and price on the y-axis. In this case, each bar represents a data point/car and the y-axis is the absolute price value for each car. 

Second, we could use the cardinality of the data attributes to define how many visual marks (e.g., bars in a bar chart) should be shown on a visualization. For example, imagine a bar chart that has manufacturers on the x-axis and price on the y-axis. In this case, we show 8 bars, each representing a manufacturer (e.g., Toyota, BMW, etc.), and the average price for each car manufacturer on the y-axis. Thus, glyphs are not representing the data points, but the cardinality of the paired data attribute. This approach would require us to have an averaged data attribute on one of the axis (e.g., average price for different manufacturers). Cardinality of data attributes that were less than 50 ranged from 5 (minimum number of visual marks) to 34 (maximum number of visual marks). In our study design, we went with this second approach.

\subsection{Tasks}


We selected the tasks for our study based on two considerations. First, tasks should be drawn from those commonly encountered while analyzing tabular data. Second, the tasks should be present in existing task taxonomies and often used in other studies to evaluate visualizations.

Previously, Amar et al.~\cite{amar2005low} proposed a set of ten low-level
analysis tasks that describe users' activities while using visualization tools
to understand their data. First, these tasks are real world tasks because users
came up with them while exploring five different datasets with different
visualization tools. Second, different studies used these tasks to evaluate
effectiveness of visualizations.
With this in mind, we used the low-level taxonomy by Amar et al.~\cite{amar2005low}, described below. 

\noindent\textbf{Find Anomalies.} We asked participants to identify any anomalies within a given set of data points with respect to a given relationship or expectation.
We crafted these anomalies manually so that, once noticed, it would be straightforward to verify that the observed value was inconsistent with what would normally be present in the data (e.g., movies with zero or negative length would be considered abnormal). For example, \textit{which genre of movies appears to have abnormal length? }

\noindent\textbf{Find Clusters.} For a given set of data points, we asked participants to count the number of groups of similar data attribute values. For example, \textit{how many different genres are shown in the chart below?}

\noindent\textbf{Find Correlation.} For a given set of two data attributes, we asked participants to determine if there is a correlation between them. To verify the responses to correlation tasks, we computed Pearson's correlation coefficient (r) to ensure that there was a strong correlation ($r \leq -0.7\text{ or }r\geq 0.7$) between the two data attributes. For example, \textit{is there a strong correlation between average budget and movie rating?}

\noindent\textbf{Compute Derived Value.} For a given set of data points, we asked participants to compute an aggregate value of those data points. For example, \textit{what is the sum of the budget for the action and the sci-fi movies?}

\noindent\textbf{Characterize Distribution.} For a given set of data points and an attribute of interest, we asked participants to identify the distribution of that attribute's values over the set. For example, \textit{what percentage of the movie genres have an average gross value higher than 10 million? }

\noindent\textbf{Find Extremum.} For this task, we asked participants to find data points having an extreme value of an data attribute. For example,\textit{ what is the car with highest cylinders? }

\noindent\textbf{Filter.} For given concrete conditions on data attribute values, we asked participants to find data points satisfying those conditions. For example, \textit{which car types have city miles per gallon ranging from 25 to 56?}

\noindent\textbf{Order.} For a given set of data points, we asked participants to rank them according to a specific ordinal metric. For example, \textit{which of the following options contains the correct sequence of movie genres, if you were to put them in order from largest average gross value to lowest?}

\noindent\textbf{Determine Range.} For a given set of data points and an attribute of
interest, we asked participants to find the span of values within the set. For example, \textit{what is the range of car prices? }

\noindent\textbf{Retrieve Value.} For this task, we asked participants to identify values of attributes for given data points. For example, \textit{what is the value of horsepower for the cars?} 

\subsection{Visualization Design}

%

To generate visualizations, we used three pairwise combinations of three different data attribute types available in our datasets. In particular, we used Nominal * Numerical, Ordinal * Numerical, Numerical * Numerical. We did not include Nominal * Nominal because it is not possible to represent this combination using all five visualizations considered in this study (e.g., line chart). 




 \begin{figure}
 \centering
  \subfloat[Retrieve Value Task]{
  \includegraphics[width=1\columnwidth]{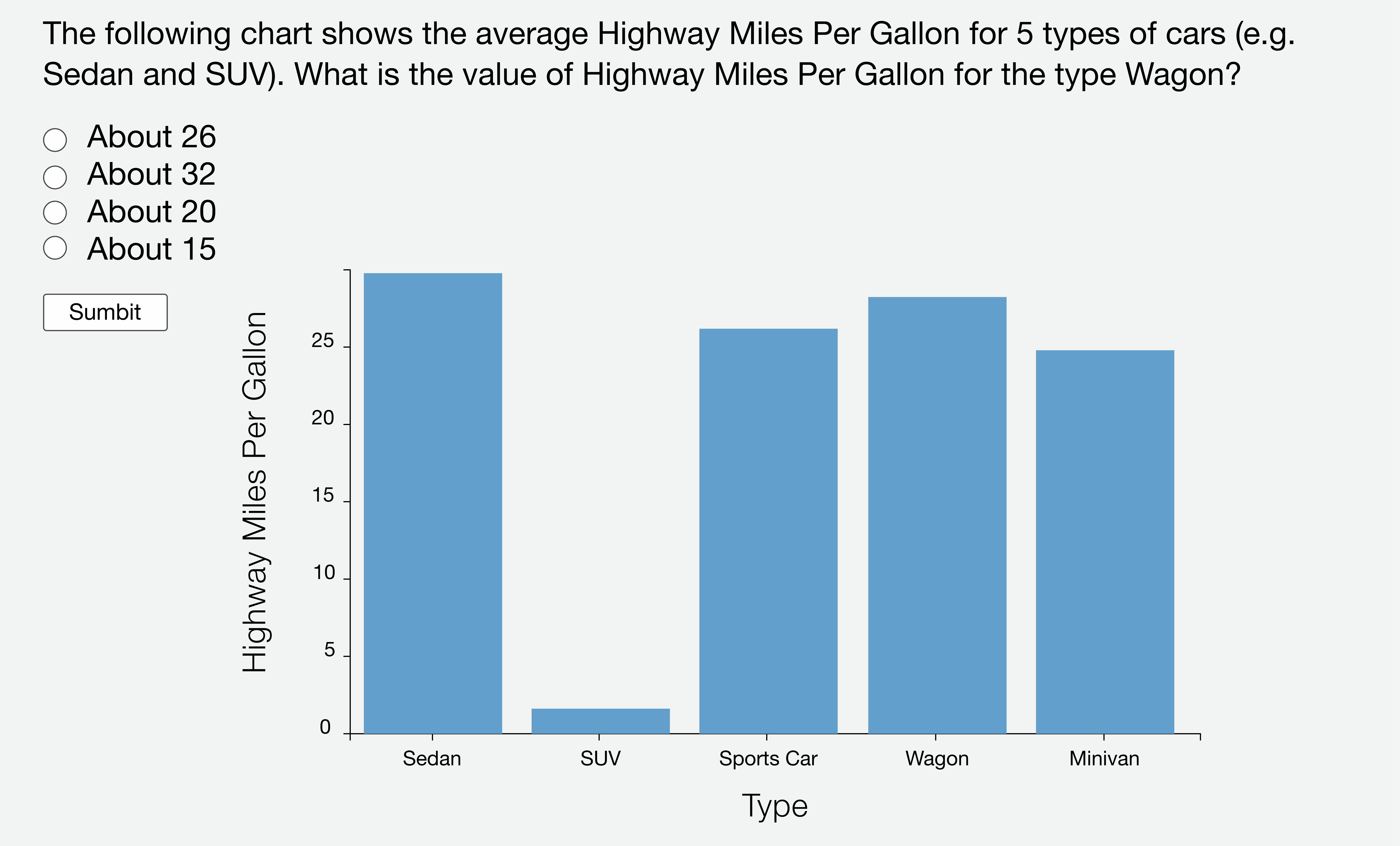}} \hfill
  \subfloat[Determine Range Task]{
  \includegraphics[width=1\columnwidth]{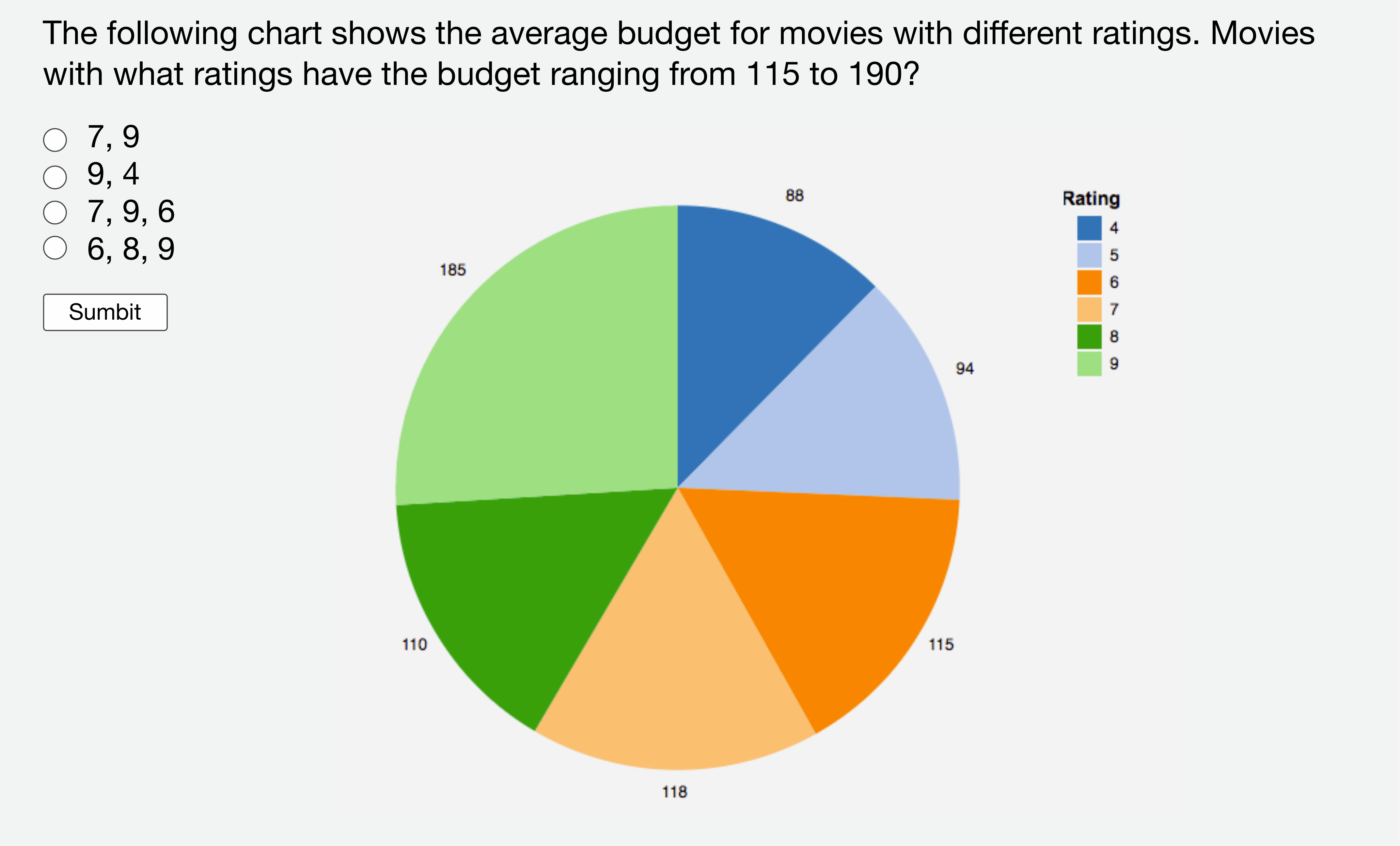}}\hfill
  \subfloat[Find Extremum]{
  \includegraphics[width=1\columnwidth]{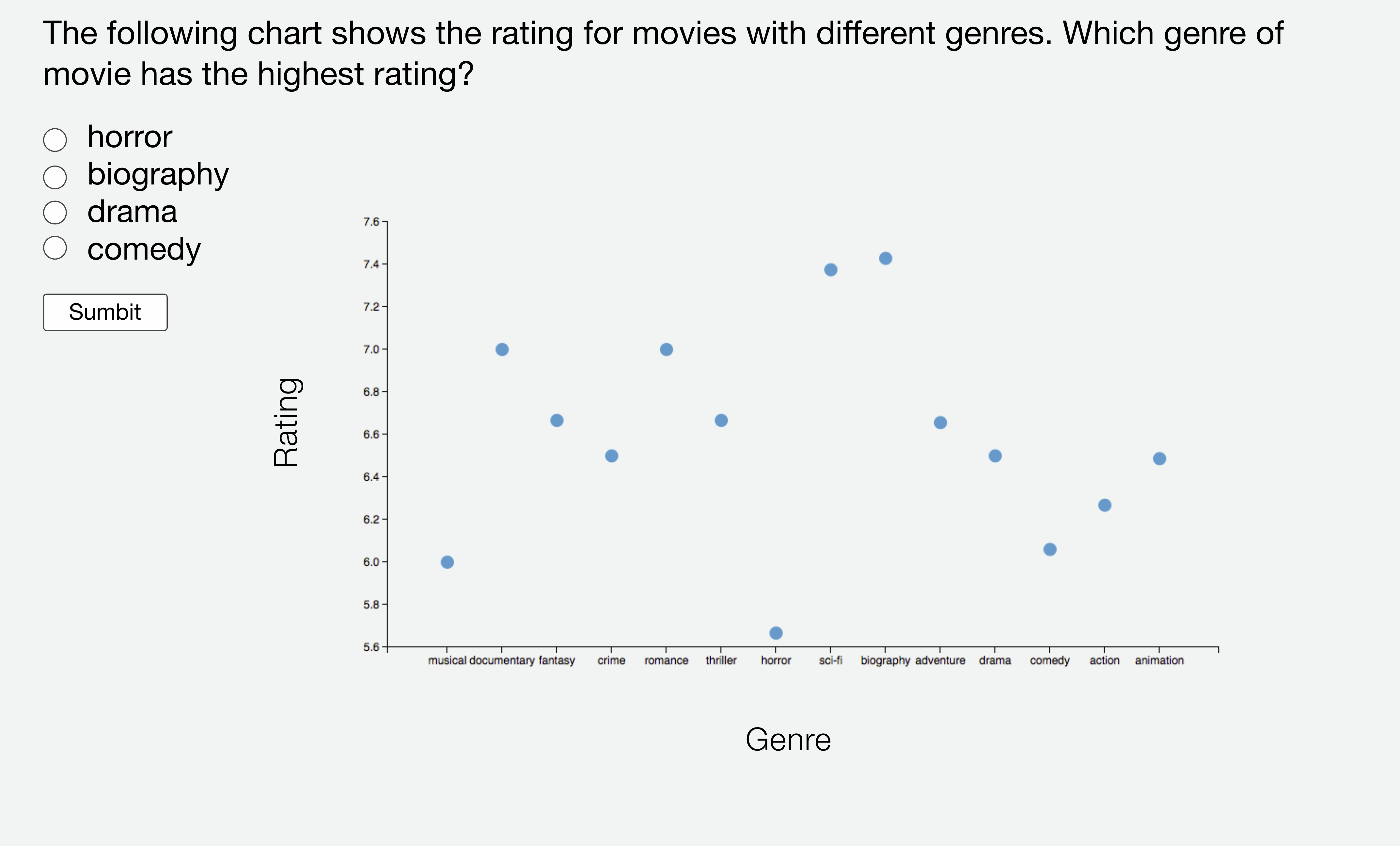}}
  
  \caption{Screenshots of three of the trials used in this experiment. Each of the trials asks users to perform a specific task.}~\label{fig:screenshot} 
\end{figure}

To create Scatterplots, Bar Charts, and Line Charts, we used the same length, font
size, and color to draw their $x-y$ axes. In addition, all the visual elements
(e.g., bars in a bar chart) used in the three charts had the same blue
color. Unlike other visualizations, pie charts do not have any axis to read the values
from. That is, to create Pie Charts we had to make design decisions on how to
show values of two data attributes used to generate them. The main design
decision that we had to make for Pie Charts was whether to include legends.
Instead of having legends, we could potentially add labels on the top of slices
of Pie Charts. We tried to put the labels on the top of slices but this caused
visual clutter, particularly in cases where the labels were long.
Additionally, using legends for Pie Charts is a common practice in
majority of commercial visualization dashboards~\cite{SpotFire, Tableau}. We decided to not show any value on the top of the slices of Pie Charts, instead showing the values of one data attribute using a legend and another one
beside the slices. For Tables, we separated different rows of the table using light gray lines. We
used a darker background color to make the labels (two data attributes used for
creating the table) distinguishable. See Figure~\ref{fig:vis} for more details.



\section{User Experiment}
In this section, we explain the details of the experiment. We make all the relevant materials for our analysis publicly available\footnote{\url{https://github.com/gtvalab/ChartsEffectiveness}}.

 \begin{figure*}
 \centering
  \includegraphics[width=\linewidth]{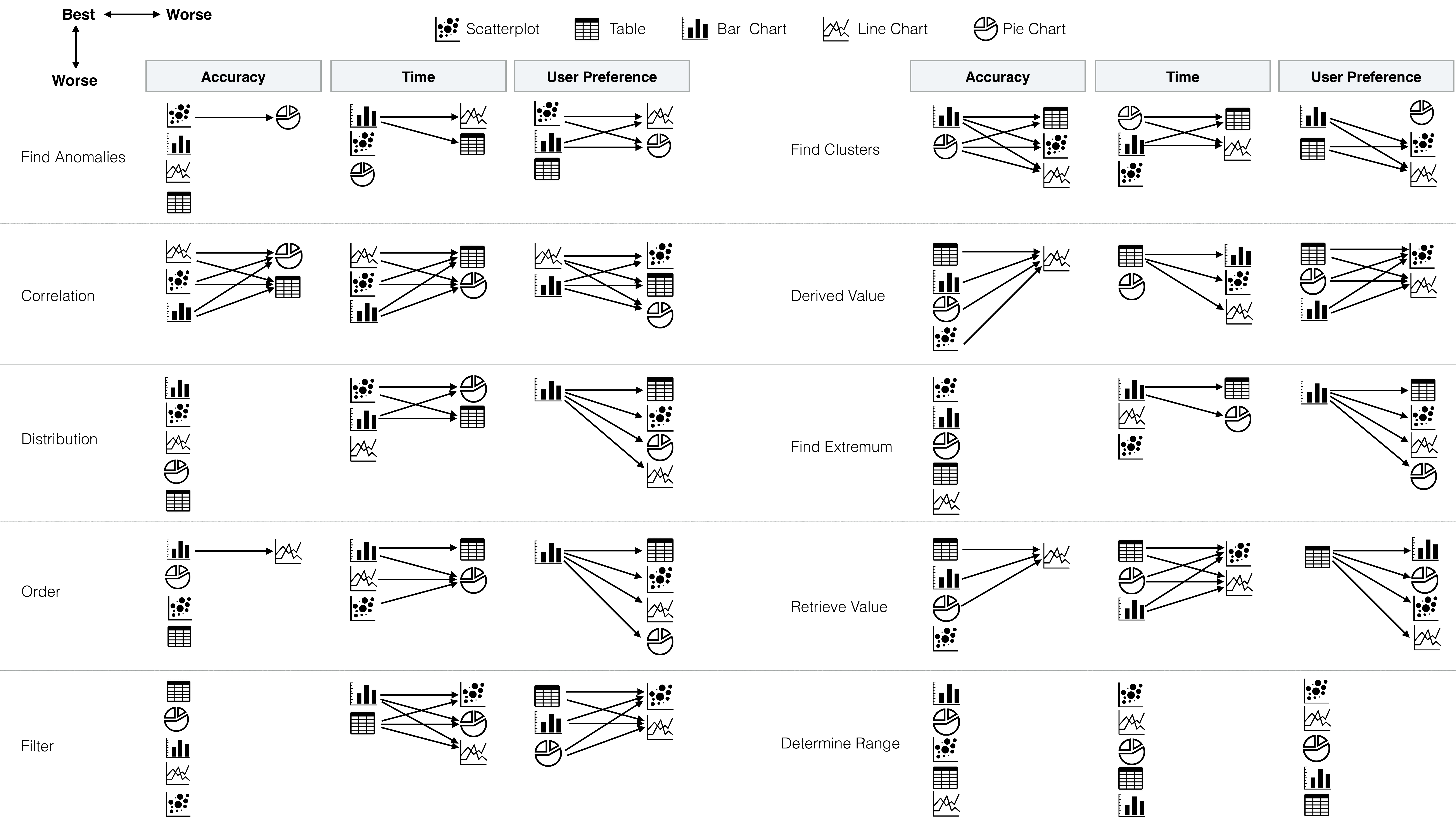}
  
  \caption{Pairwise relation between visualization types across tasks and performance metrics. Arrows show that the source is significantly better than the target. }~\label{fig:summaryFig}
\end{figure*}

\subsection{Experimental Platform \& Participants}

We conducted our experiment by posting it as a job, Human Intelligence Task
(HIT), on Amazon's Mechanical Turk (MTurk). To be able to participate in our study, MTurk workers (who perform tasks posted on MTurk), had to have an approval rate of 95\% and at least 100 approved HITs as a quality check.  We implemented our experiment as a web application hosted on a server external to MTurk. Participants accessed the experiment through a URL link  posted on the MTurk site. Each worker could participate in our study only once. The study took about 25 to 40 minutes to complete and we compensated the workers who participated \$4.

In order to determine the minimum number of participants needed for our
study, we first conducted a pilot study with 50 participants on Amazon's
Mechanical Turk. Based on the data collected from our pilot study, we conducted
a statistical power analysis to ensure that our experiment included enough
participants to reliably detect meaningful performance differences across
independent variables of the experiment. Our power analysis, based on the results of the pilot study, indicated that at least 160 participants would be required to detect a large effect.

After determining the number of subjects required to participate in our study,
we recruited 203 workers to participate in our study. Among the 203 who participated in our study 180 of them (105 Male, 75 Female) completed the study. The age of our workers ranged
from 25--40 years. All workers participated in our experiment were based in the
United States and have used visualizations before. 107 of the participants had
experience creating visualizations using Microsoft Excel. Five of the participants 
also had experience in creating visualizations using Tableau software.

\subsection{Procedure}

\noindent\textbf{Training.} Before starting the main experiment, participants were briefed
about the purpose of the study and their rights. At this stage, the
participants were also asked to answer to some demographic questions 
(e.g., age, sex, and prior experience in creating visualizations). Participants were then asked to perform 5 trial questions (one question per visualization) as quickly and accurately as possible. Trial questions were presented in a random order. For each participant, the training questions were a randomly ordered set of these five questions. During this session, after answering each question participants received feedback that showed the correctness of their answers. To prevent the participants from skipping the training questions, participants were not able to move to the next question unless they answered the question correctly. 


\noindent\textbf{Main Experiment.} During the main experiment 180 participants were randomly assigned to 10 tasks (18 participants per task). So, each participant performed questions designed for one type of task. For each type of task, we had 30 questions $( 5 \ Visualizations \times \ 2 \ Datasets \times \  3 \ Trials)$. As recommended by previous work~\cite{olson2014ways}, we also designed two additional questions to detect if a participant answered the questions randomly. These two questions were straightforward and designed to make sure that participants read the questions. Questions were presented in a random order to prevent participants from extrapolating new judgments from previous ones. We counterbalanced the difficulty (number of visual marks shown in a visualization) of the questions for each visualization type. Screenshots of the questions for the main experiment are shown in Figure~\ref{fig:screenshot}. More task screenshots are provided in our supplemental materials\footnote{\url{https://github.com/gtvalab/ChartsEffectiveness}}).

\noindent\textbf{Follow Up Questions.} After completing the main experiment, the participants were asked to
perform 6 additional ranking questions $(3 \ Trials \times \ 2 \ Datasets)$. In each ranking question the participants
were asked to rank the five different visualizations in the order of their
preference for performing this task. 
Before finishing the experiment, we asked participants to \textit{"Please enter the criteria you used for ranking the charts along with any other additional comments you have about the experiment in general"}. This was to allow the participants to convey their feedback and in order to solicit potentially unexpected insights.

Questions (training questions, main experiment questions, and ranking questions) were pre-generated in an iterative process by all three authors in multiple sessions. After each session, we conducted a pilot study to extract the major problems with the designed questions. We had two criteria while designing questions for our experiment. First, for our study to have a reasonable length and complexity, we had to design questions with a reasonable level of difficulty. For example, questions with a high level of difficulty could frustrate the participants. To define difficulty, we designed the questions in a way that the average response time for a single question is in the range from 10 to 40 seconds. Second, questions are balanced across different datasets and presented comparable values. For example, if a categorical attribute in the movies dataset had five categories, we tried to pick a variable from the cars dataset that also had five (or around five) categories.

\subsection{Data Analysis}


To analyze the differences among the various visualizations, for each participant, we calculated mean performance values for each task and visualization type. That is, we averaged time and accuracy of questions for each visualization type and task. Before testing, we checked that the collected data met the assumptions of appropriate statistical tests. The assumption of normality was not satisfied for performance time. However, the normality was satisfied for log transformation of time values. So, we treated log-transformed values as our time measurements. We conducted repeated-measures analysis of variance (ANOVA) for each task independently to test for differences among the
various visualizations, datasets, and their interactions with one another. While the Visualization had significant effects on both accuracy and time, the Dataset had no significant effect on accuracy or time.

\begin{table*}[p] \scriptsize
       \caption{This figure shows performance results for 10 different tasks. Performance results for each task are shown using three sub-charts. Mean accuracy results are shown on the left (mean accuracy is measured in percentage), mean time results are shown in the middle, and user preferences/rankings are shown at the right (1 shows least preferred and 5 shows the most preferred). Statistical test results are also shown below the charts. All tests display 95\% confidence intervals and are Bonferroni-corrected.} ~\label{TAB:taskFig1} 
    \begin{minipage}{0.5\linewidth}
        \begin{tabular}{p{8.7cm}}
        Find Anomalies \\
        \midrule
	      \includegraphics[width=\linewidth, keepaspectratio]{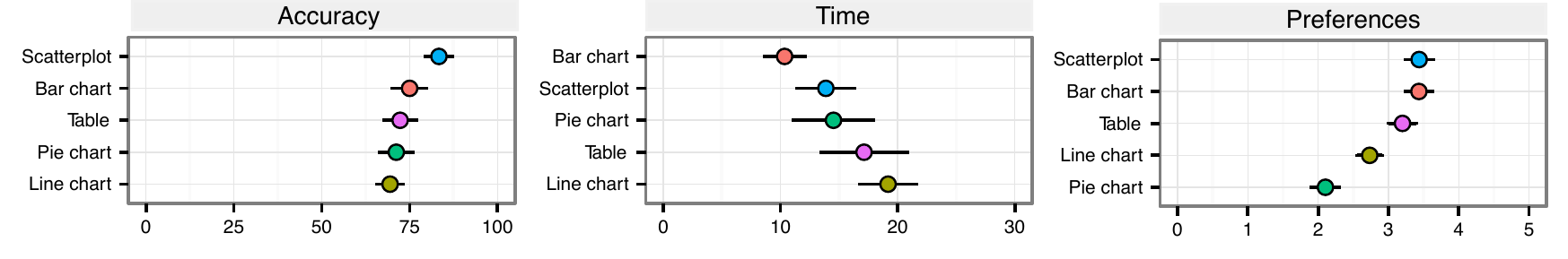}\\\\
             
             \textbf{Accuracy:} ($F_{(3.4,4915.1)} = 3.03, p < 0.05$, $\eta_{p}^{2} = 0.15$) \\
             Results of Bonferroni-corrected post-hoc comparisons showed that Line Chart was significantly less accurate than Scatterplot ($p <0.05$).\\ \\
             
             \textbf{Time:} ($F_{(4,68)} = 0.48, p < 0.05$, $\eta_{p}^{2} = 0.27$)\\
             Posthoc comparisons indicate that Bar Chart was significantly faster than Line Chart and Table ($p <0.05$). This might be because people can decode values encoded with length faster than other encodings such as angle or distance~\cite{cleveland1984graphical, simkin1987information, Talbot_bar}.\\\\
             
            \textbf{Preference:} ($F_{(3.1,45.56)} = 5.9, p< 0.05$, $\eta_{p}^{2} = 0.26$)\\
            For the Anomalies task type, results of pairwise comparisons show that user preference in performing Anomalies tasks using Bar Chart and Scatterplot were significantly higher than Pie Chart and Line Chart ($p <0.05$). \newline \newline
            
  \\
        \end{tabular}
    \end{minipage} 
  \begin{minipage}{0.5\linewidth}
        \begin{tabular}{p{8.7cm}}
        Find Clusters \\
        \midrule
	      \includegraphics[width=\linewidth, keepaspectratio]{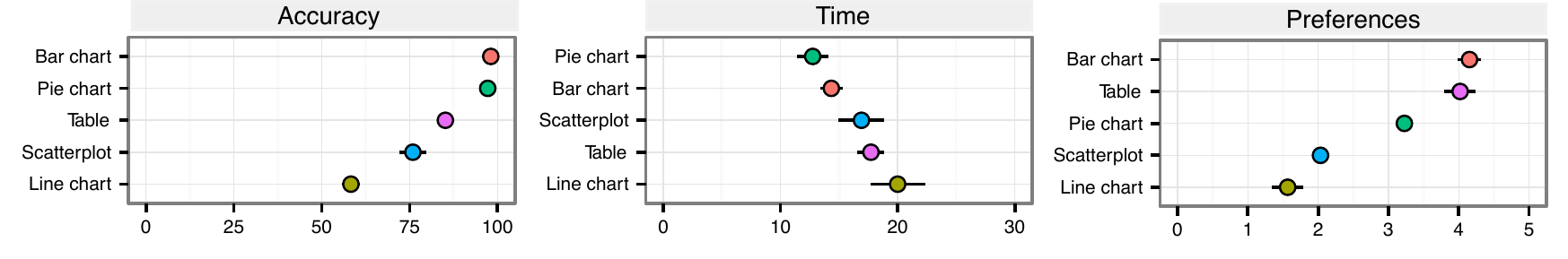}\\\\
             
             \textbf{Accuracy:} ($F_{(2.6,45065.1)} = 60.7, p < 0.05$, $\eta_{p}^{2} = 0.78$)\\
             Results of Bonferroni-corrected posthoc comparisons show that Pie Chart and Bar Chart were significantly more accurate than other visualizations. ($p <0.05$).\\ \\
             
             \textbf{Time:} ($F_{(3.9,67.9)} = 6.9, p < 0.05$, $\eta_{p}^{2} = 0.29$) \\ 
              Pie Chart and Bar Chart were significantly faster than Table ($p <0.05$) and line Chart ($p <0.05$. We believe that uniquely coloring different slices of pie charts improved the performance of Pie Chart for this type of tasks.\\\\
             
            \textbf{Preference:} ($F_{(2.9,188.56)}
= 30.2, p < 0.05$, $\eta_{p}^{2} = 0.64$)\\
User preferences in using Bar Chart and Table were significantly higher than other visualizations ($p <0.05$). While user preferences in using Bar Chart can be explained by its high accuracy and speed, it is surprising that Table was also highly preferred by users for Cluster tasks.
	    \\ \\
        \end{tabular}
    \end{minipage}

      \begin{minipage}{0.5\linewidth}
        \begin{tabular}{p{8.7cm}}
        Correlation \\
     \midrule
	      \includegraphics[width=\linewidth, keepaspectratio]{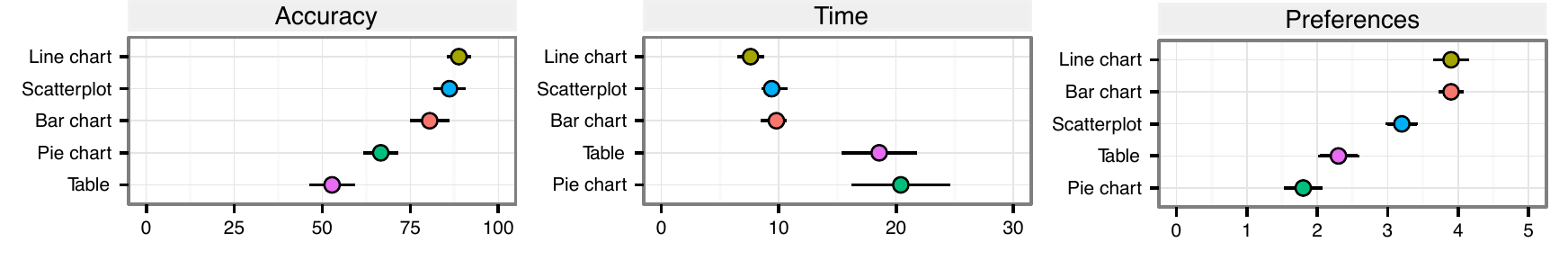}\\\\
             
             \textbf{Accuracy:} ($F_{(2.5, 20528.2)} = 12.1, p < 0.05$, $\eta_{p}^{2} = 0.41$)  \\
             Pairwise comparison show that Line Chart and Scatterplot were significantly more accurate than other charts ($p<0.05$). Bar Chart was also significantly more accurate than Pie Chart and Table ($p<0.05$).\\ \\
             
             \textbf{Time:} ($F_{(1, 479.7)} = 42.3, p < 0.05$, $\eta_{p}^{2} = 0.7$) \\
             We found that Line Chart, Bar Chart and Scatterplot were significantly faster than Pie Chart and Table ($p<0.05$). In fact, our results validates the findings of the previous work that showed the effectiveness of Scatterplots and Line charts for Correlation tasks~\cite{harrison2014ranking, Pandey:2016}. \\\\
             
            \textbf{Preference:} ($F_{(3.6, 75.2)} = 13.6, p
< 0.05$, $\eta_{p}^{2} = 0.44$)\\
            User preference in performing Correlations tasks using Bar Chart and Line Chart were significantly higher than that of Pie Chart, Scatterplot, and Table ($p <0.05$).   \newline 
	    \\ \\ 
        \end{tabular}
    \end{minipage} 
          \begin{minipage}{0.5\linewidth}
        \begin{tabular}{p{8.7cm}}
        Compute Derived Value \\
       \midrule
	      \includegraphics[width=\linewidth, keepaspectratio]{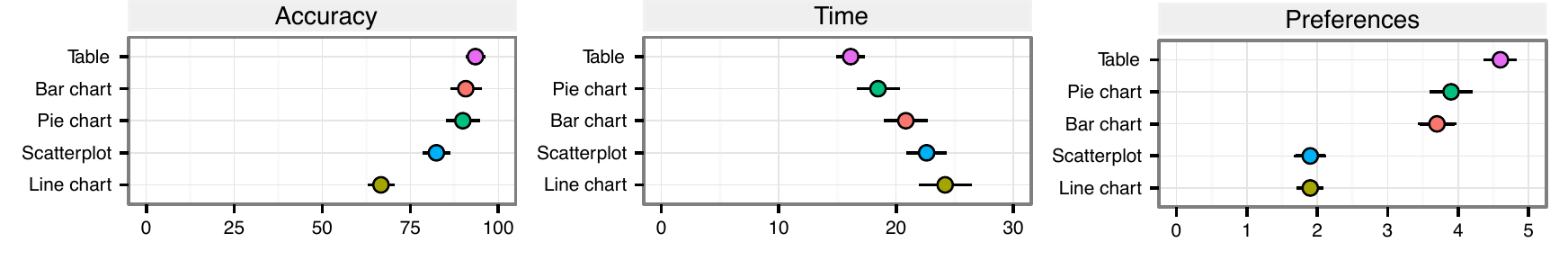}\\\\
             
             \textbf{Accuracy:} ($F_{(2.7, 18234.2)} = 16.2, p < 0.05$, $\eta_{p}^{2} = 0.49$)\\ 
             Accuracy of Line Chart was significantly lower than rest of the four chart types ($p <0.05$). On the other hand, there was no significant difference among Bar Chart, Scatterplot, Pie Chart, and Table. High accuracy of Pie Chart may have been further helped by having text labels showing the data values.\\ \\
             
             \textbf{Time:} ($F_{(3.2, 0.4)} = 9.6, p < 0.05$, $\eta_{p}^{2} = 0.36$)\\ 
            Table was significantly faster than Bar Chart ($p <0.05$), Scatterplot ($p <0.05$), and Line Chart ($p <0.05$) for this type of tasks. High effectiveness of Table might be because the exact values for each data point is shown in tables. So it might be the case that less cognitive work is required to aggregate the values when the exact values are shown.\\\\
             
            \textbf{Preference:}  ($F_{(3.1, 187.8)} = 35.3,
p < 0.05$, $\eta_{p}^{2} = 0.67$)\\
            Participants preference for using Table, Pie Chart, and Bar Chart is significantly higher than Scatterplot ($p <0.05$) and Line Chart ($p <0.05$). 
\\ \\ \\ 
        \end{tabular}
    \end{minipage}

     \begin{minipage}{0.5\linewidth}
        \begin{tabular}{p{8.7cm}}
            Characterize Distribution \\
        \midrule
	      \includegraphics[width=\linewidth, keepaspectratio]{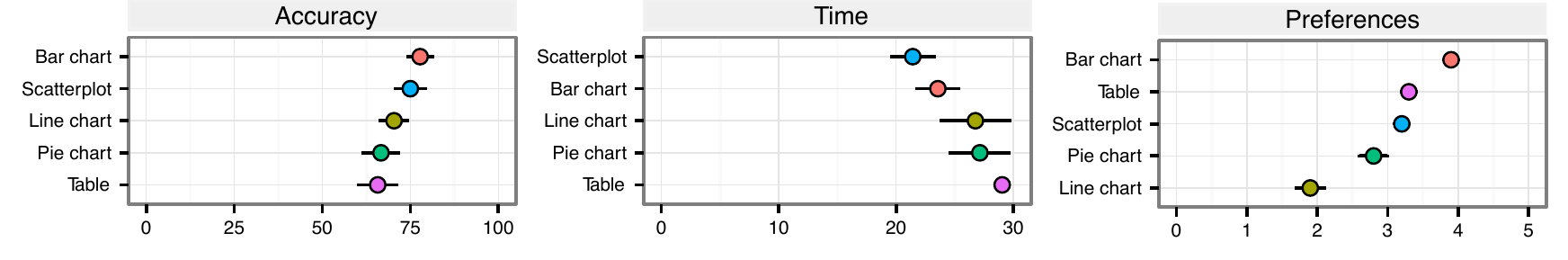}\\\\
             
             \textbf{Accuracy:} No significant main effect was found.
            \\ \\
             
             \textbf{Time:} ($F_{(4, 68)} = 5.6, p < 0.05$, $\eta_{p}^{2} = 0.25$)  \\
             Our results indicate that Scatterplot and Bar Chart are significantly faster than Pie Chart ($p <0.05$) and Table ($p <0.05$) for Distribution tasks. Previous work also showed the fast speed of Scatterplot for correlation tasks~\cite{harrison2014ranking, 2016-beyond-webers-law}. \\\\
             
            \textbf{Preference:} ($F_{(2.5, 20528.2)} = 12.1, p < 0.05$, $\eta_{p}^{2} = 0.41$)\\
            Our results indicate that participants preferred Bar Chart, Scatterplot, and Table significantly more than Pie Chart ($p <0.05$) and Line Chart ($p <0.05$). It is surprising that even though Table was not faster than the other four visualizations, participants highly preferred using it.   
        \end{tabular}
    \end{minipage} 
          \begin{minipage}{0.5\linewidth}
        \begin{tabular}{p{8.7cm}}
         Find Extremum\\
       \midrule
	      \includegraphics[width=\linewidth, keepaspectratio]{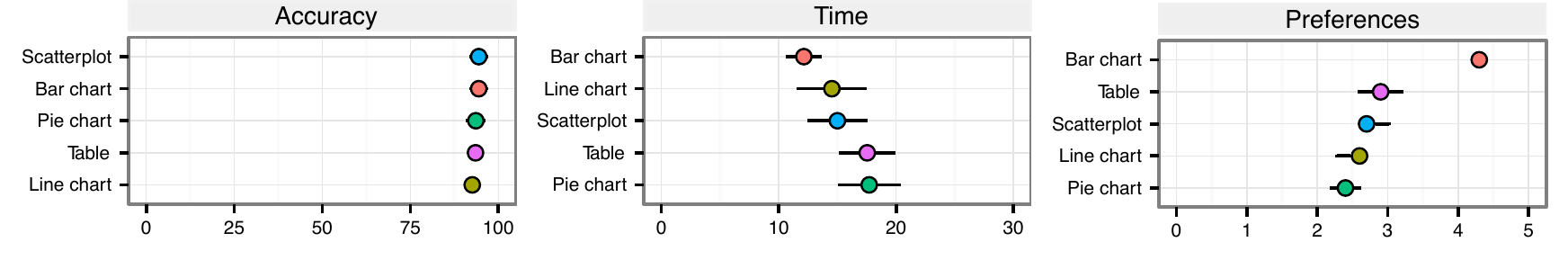}\\\\
             
             \textbf{Accuracy:} No significant main effect was found. \\ \\
             
             \textbf{Time:} ($F_{(4, 0.4)} = 10.4, p < 0.05$, $\eta_{p}^{2} = 0.38$)\\ 
             
             Bar Chart is significantly faster than Table ($p <0.05$) and Pie Chart ($p <0.05$). Previous work also recommends using Bar Chart in cases where readers are looking for a maximum or minimum values~\cite{few2006information}. \\\\
             
            \textbf{Preference:} ($F_{(2.8, 89.4)} = 8.2, p < 0.05$, $\eta_{p}^{2} = 0.61$)\\
            There is a significant main effect of Visualization on user preference. For Extremum tasks, participants' preference in using bar charts is significantly higher than all other visualizations ($p <0.05$).\newline 
        \end{tabular}
    \end{minipage} 
\end{table*}

\begin{table*} \scriptsize
    \begin{minipage}{0.5\linewidth}
        \begin{tabular}{p{8.7cm}}
           Order \\
       \midrule
	      \includegraphics[width=\linewidth, keepaspectratio]{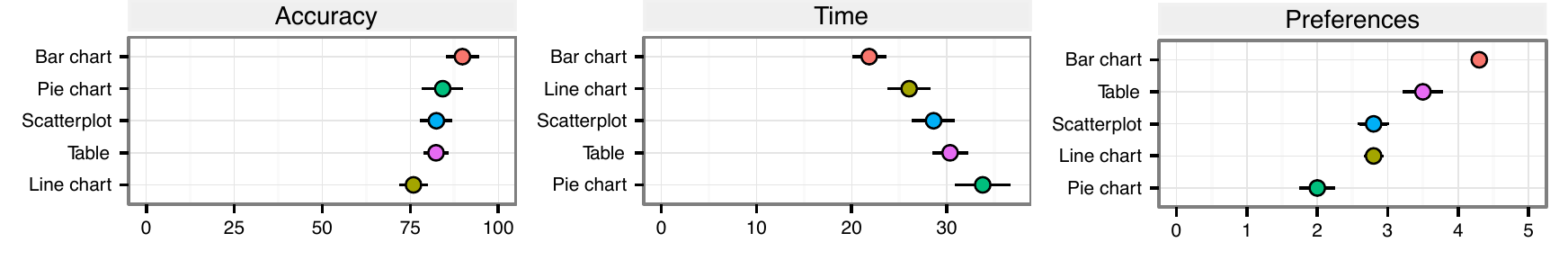}\\\\
             
             \textbf{Accuracy:} ($F_{(4, .03)} = 2.6, p < 0.05$, $\eta_{p}^{2} = 0.17$) \\
             Bar Chart is significantly more accurate than Line Chart ($p <0.05$). We did not find a significant difference among Bar Chart, Pie Chart, Scatterplot, and Table.  \\ \\
             
             \textbf{Time:} ($F_{(3.3, 0.6)} = 9.3, p < 0.05$, $\eta_{p}^{2} = 0.35$)\\
            Bar Chart is significantly faster than Pie Chart ($p<0.05$) and Table  ($p<0.001$). Line Chart is also significantly faster than Table  ($p<0.05$) for Order tasks. We also found that Scatterplot is significantly faster than Pie Chart ($p<0.05$). High performance of Line Chart, Scatterplot, and Bar Chart could be the due to their usage of length and position as primary graphical encodings. Length and position are fastest encodings to perceive~\cite{cleveland1985graphical, cleveland1984graphical}.\\\\
             
            \textbf{Preference:} ($F_{(3.0, 103.3)} = 11.8, p < 0.05$, $\eta_{p}^{2} = 0.52$)\\
            For Order tasks, users preferred Bar Chart significantly more than other visualizations ($p <0.05$). Moreover, our results indicate that user preference in using Pie Chart is significantly lower than other visualizations. There was not a significant different in user preference for Line Chart and Scatterplot.
\\  \\ \\ \\
        \end{tabular}
    \end{minipage} 
  \begin{minipage}{0.5\linewidth}
        \begin{tabular}{p{8.7cm}}
            Retrieve Value\\
       \midrule
	      \includegraphics[width=\linewidth, keepaspectratio]{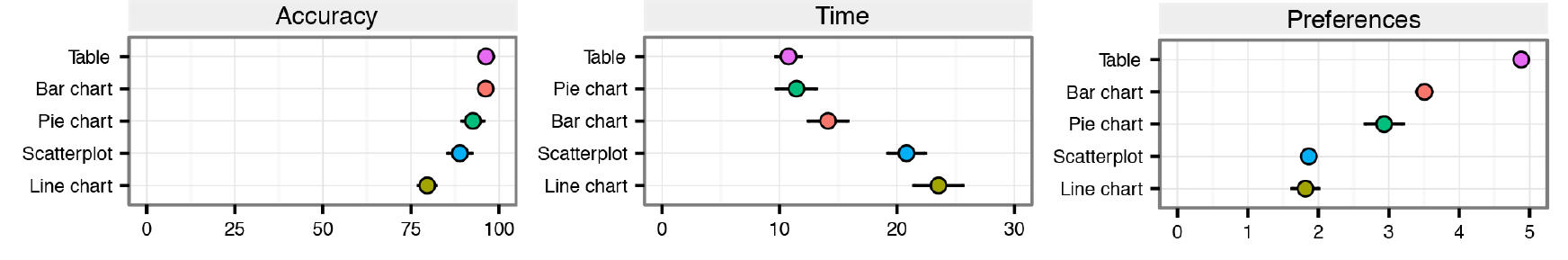}\\\\
             
             \textbf{Accuracy:} ($F_{(2.9,7114.1)} = 7.7, p < 0.05$, $\eta_{p}^{2} = 0.32$)\\
             Overall, Bar Chart, Table and Pie Chart were significantly more accurate than Line Chart ($p <0.05$). The difference between accuracy in Scatterplot and Line Chart was not significant. We would like to mention that Pie Chart may have been further helped by having text labels showing the data values.\\ \\
             
             \textbf{Time:} ($F_{(3.0,52.1)} = 4.34, p < 0.05$, $\eta_{p}^{2} = 0.26$) \\ 
             Table, Pie Chart, and Bar Chart are significantly faster than Scatterplot ($p<0.05$) and Line Chart ($p<0.05$) for performing Retrieve tasks. Successful performance time of Retrieve tasks highly depends on readers ability to rapidly identify the value for a certain data point. As Ehrenberg~\cite{ehrenberg1975data} points out, tables are well-suited for retrieving the numerical value of a data point when a relatively small number of data points are displayed.\\\\
             
            \textbf{Preference:} ($F_{(1.5, 417.2)} = 47.1, p < 0.05$, $\eta_{p}^{2} = 0.73$) \\
             User preference for performing Retrieve tasks using Table is significantly higher than other visualizations. After Table, Bar Chart is the second most visualization type highly preferred by users to perform this type of tasks ($p <0.05$). Moreover, user preference in using Bar Chart is significantly higher than Pie Chart, Scatterplot, and Line Chart.

        \end{tabular}
    \end{minipage}

      \begin{minipage}{0.5\linewidth}
        \begin{tabular}{p{8.7cm}}
             Filter\\
       \midrule
	      \includegraphics[width=\linewidth, keepaspectratio]{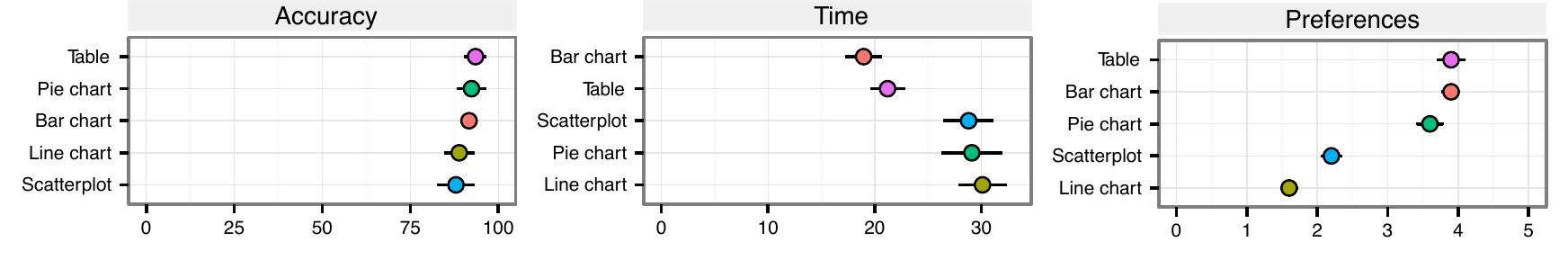}\\\\
             
             \textbf{Accuracy:} No significant main effect was found. \\ \\
             
             \textbf{Time:} ($F_{(2.2, 210.5)} = 42.2, p < 0.05$, $\eta_{p}^{2} = 0.72$) \\
             Bar Chart and Table are significantly faster than other visualizations ($p<0.05$). \\\\
             
            \textbf{Preference:} ($F_{(3.6, 75.2)} = 13.6, p
< 0.05$, $\eta_{p}^{2} = 0.44$) \\
            Participants' preference towards using Table, Bar Chart, and Pie Chart is significantly higher than Line Chart ($p<0.05$) and Scatterplot ($p<0.05$) for Filter tasks. 
        \end{tabular}
    \end{minipage} 
          \begin{minipage}{0.5\linewidth}
        \begin{tabular}{p{8.7cm}}
                 Determine Range\\
       \midrule
	      \includegraphics[width=\linewidth, keepaspectratio]{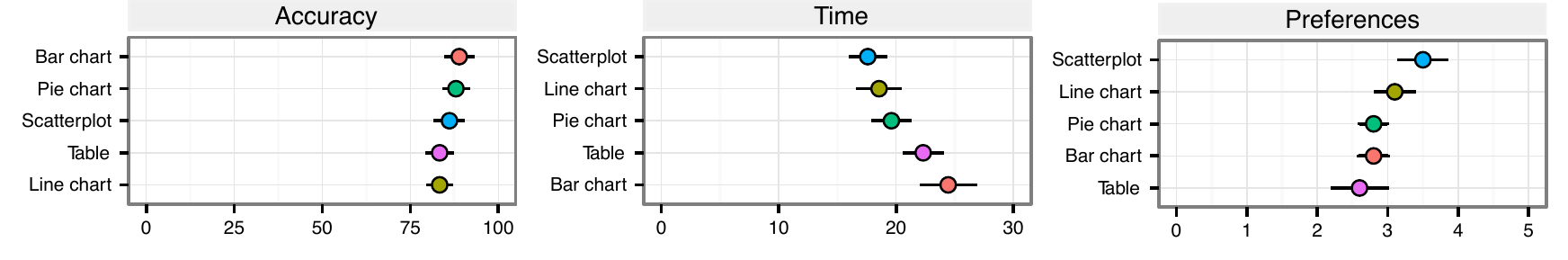}\\\\
             
             \textbf{Accuracy:} No significant main effect was found.
             \\ \\ \\ 
             
             \textbf{Time:}
            No significant main effect was found.  \\ \\ \\
             
            \textbf{Preference:} No significant main effect was found.\\ \\

        \end{tabular}
    \end{minipage} 
\end{table*}

\section{Results}

We first give an overview of our analysis of the results and then discuss them in detail for each task. We provide  detailed  analysis  of  the  results in Table~\ref{TAB:taskFig1}. Throughout the following sections, accuracy refers to values in percentages (\%) and time refers to values in seconds. 

Results, aggregated over tasks and datasets, show that Bar Chart is the fastest and the most accurate visualization type. This result is in line with prior work on graphical perception showing that people can decode values encoded with length faster than other encodings such as angle or volume~\cite{cleveland1984graphical, simkin1987information, Talbot_bar}. Conversely, Line Chart has the lowest aggregate accuracy and speed. However, Line Chart is significantly more accurate than other charts for Correlation and Distribution tasks. This finding concurs with earlier research reporting the effectiveness of line charts for trend finding tasks (e.g.,
~\cite{zacks1999bars}). Nonetheless, the overall low performance of Line Chart is surprising and, for some tasks, can be attributed to the fact that the axes values ("ticks") were drawn at
intervals. This makes it difficult to precisely identify the value for a specific data point.

While Pie Chart is comparably as accurate and fast as Bar Chart and Table  
for Retrieve, Range, Order, Filter, Extremum, Derived and Cluster tasks, it is 
less accurate for Correlation, Anomalies and Distribution tasks. 
Pie Chart is the fastest visualization for performing Cluster task. 
The high performance of Pie Chart for these tasks can be attributed to its relative effectiveness in conveying part-whole relations and facilitating 
proportional judgments, particularly when the number of data points 
visualized is small~\cite{eells1926relative, spence1991displaying}. 
Pie Chart may have been further helped by having colored slices with
text labels showing the data values.  

Overall, Scatterplot performs reasonably well in terms of both accuracy and time. For the majority of tasks Scatterplot is among the most effective top three
visualizations, and it was never the least accurate or slowest visualization
for any of the tasks.


Bar Chart and Table are the two visualization types highly preferred by participants across most of the tasks. Bar Chart is always among the two top-performing visualizations for almost all tasks, so this makes sense that people prefer using Bar Chart over other visualizations. Surprisingly, while performing some of the tasks (e.g., Distribution, Anomalies) using Table is relatively slow and less accurate, participants still prefer Table for performing these tasks. People's familiarity with tables and ease of understanding tables could have led people to prefer using tables over other visualizations. To determine whether performance time and accuracy are related to user
preferences, we calculated the correlation between performance time, accuracy,
and user preference. We found a positive correlation between accuracy and user
preference (Pearson's $r_{(5)}= 0.68, p < 0.05$), indicating people have a preference for visualizations that allow them to accurately complete a task. We also found a weak negative correlation between performance time and user preferences (Pearson's $r_{(5)}= -0.43, p < 0.05$). 

\section{Discussion}
In this section, we reflect on the results of our work more broadly with respect to information visualization.

\subsection{No One Size Fits All}
Depending on the task at hand, various visualizations perform
differently. That is, we do not advocate generalizing the
performance of a specific visualization on a particular task to every task.
For example, throughout the history of the graphical perception research, 
pie charts have been the subject of passionate arguments~\cite{eells1926relative, 
croxton1927bar, spence1991displaying} for and against their use.
Although the current common wisdom among visualization researchers is to avoid them, 
pie charts continue to be popular in everyday visualizations. Results of our study 
present a more nuanced view of pie charts. We found that pie charts can be as effective
as other visualizations for task types such as Cluster, Extremum, Filter, Retrieve, and Range. 
On the other hand, our results suggest that pie charts perform poorly 
in Correlation and Distribution tasks.  



\subsection{User Preferences}

Our results show user preferences correlate with user accuracy and speed in completing tasks. Before completing the study, we asked participants to explain the criteria they used for ranking the visualizations. Some of participants explicitly mentioned perceived accuracy of the charts as one of the factors that influenced their decision while ranking visualizations. For example, one of the participants stated: \textit{``Just by how accurate I felt my own answer was, and how easy it was to derive the answer from the graphs.''} 

Neither accuracy nor speed appear to be the only criteria by which participants describe their individual rankings. 
Additionally, perceived accuracy does not always match with task accuracy. We noticed that for some task types such as Distribution and Cluster, preference for using tables and bar charts is significantly higher than other visualizations, even though these two visualizations are not the most effective ones for these type of tasks. Interestingly, some of the participants took into account their familiarity with visualizations as one of the factors for preferring some visualization over others. For example, one of the participants mentioned: \textit{``I just went with the ones I felt were familiar to me.''}  Another participant also stated: \textit{``I deal with bars a lot. I know how to read them.''}

\subsection{Which Visualization Type to Use?}
Based on our results, when time, accuracy and preference are important factors to consider, we provide the following guidelines: \\


\noindent\textbf{G1. Use bar charts for finding clusters.} Our results show that pie charts and bar charts are significantly faster and more accurate for this type of task. However, users preference in using bar charts was significantly higher than using pie charts for finding clusters. Thus, bar chart having a better overall performance in terms of time, accuracy, and user preferences for finding clusters.  

\noindent\textbf{G2. Use line charts for finding correlations.} We found that line charts and scatterplots have significantly higher accuracy and speed for finding correlations. However, users preferences in using line charts for finding correlations was signifincalty higher than using scatterplots. Thus, line charts performed better in terms of time, accuracy and user preferences. 

\noindent\textbf{G3. Use scatterplots for finding anomalies.} Results of
our study indicate that scatterplots have high accuracy, speed,
and are highly preferred by users for this type of task.

\noindent\textbf{G4. Avoid line charts for tasks that require readers to precisely identify the value of a specific data point.} The low performance of line charts for some tasks such as Derived Value and Cluster might be attributed to the fact that the axes values (i.e., the ``ticks'') were drawn at
uniform intervals. This makes it difficult to precisely identify the value of a specific data point. 

\noindent\textbf{G5. Avoid using tables and pie charts for correlation tasks.} Findings indicate that Tables and pie charts are significantly less accurate, slower, and less preferred by users for this type of task.

\subsection{How to engineer empirical user performance data into practical systems?}
Graphical perception experiments are the work-horse of our quest to understand and improve the effectiveness of visualizations. Guidelines and heuristics that we use today in data visualization are primarily due to the accumulation of experimental results over decades. It is not, however, always possible to extract guidelines from data collected by user experiments. Even when this is possible, such derived guidelines require visualization practitioners to manually incorporate them in visualizations systems. We believe machine learning models provide a practical opportunity to implicitly engineer the insights embodied by empirical performance data into visualization systems in an unbiased and rigorous manner. Kopol is a basic example of how this can be achieved. To drive Kopol, we train a decision tree on the  data we collected. Kopol then uses the learned model to recommend visualizations at ``test'' time for given user tasks and datasets.


\section{Limitations and Future Work}
Our experimental results should be interpreted in the context of the specified visualizations tasks, and datasets. While our findings should be interpreted in the context of the specified settings and conditions, we tested the most common visualization techniques incorporated in various visualization dashboards~\cite{LeeVlat}, analytical tasks used in different studies~\cite{amar2005low,Amar:2004} and datasets used in various studies~\cite{TableauData, UCIDataset}. That being said, additional studies are required to test our research questions taking into account different visualization techniques, tasks and datasets.


In this study, participants were required to perform the tasks using static visualizations. While we are aware of the importance of interactivity and the fact that interactivity could impact user experience with a specific visualization, we decided to exclude interactivity because of the following reasons. First, adding interactivity increases the complexity of the study design. In fact, it would require us to take into account another set of factors including users' input devices such as mouse, trackpad, and touch. Moreover, we had to take into account interaction design/implementation. For example, the implementation of each interaction varies across different input devices. Second, static visualizations are commonly used for presentation and educational purposes (e.g., visualization used in books, newspapers, and presentations). In many of these cases, visualization consumers still need to perform a variety of tasks using static visualizations. That being said, we encourage additional studies to directly investigate the effectiveness of these visualizations taking into account interactivity.

Due to practical limitations of conducting the study using static visualizations with a large number of visual marks (e.g., length and complexity of the experiment), the number of visual marks shown in the visualizations used in our study is restricted to be between 5 to 34. We used the cardinality of the data attributes to define how many visual marks (e.g., bars in a bar chart, circles in a scatterplot) should be shown in a visualization. However, we would like to emphasize that performance of these visualization types might change depending on the number of data points encoded  by them.  Our study results hold for static visualizations with visual marks that number between 5 and 34. We defer investigation of how datapoint cardinality affects the task-based performance of visualizations to future work. 

In this study, we investigated the effectiveness of five basic two-dimensional visualization types. However, some of the visualization types can be extended to more than two dimensions (e.g., line chart). Performance of these visualization types might change depending on their dimensionalities. One interesting avenue of continued research is to investigate the impact of the number of dimensions represented by a visualization type on its effectiveness.

\section{Conclusion}

In this work, we report the results of a study that gathers user performance and preference for performing ten common data analysis tasks using five small scale (5-34 data points) two-dimensional visualization types: Table, Line Chart, Bar Chart, Scatterplot, and Pie Chart. We use two different datasets to further support the ecological validity of results. We find that the effectiveness of the visualization types considered significantly changes from one task to another. We compile our findings into a set of recommendations to inform data visualization in practice.

\bibliographystyle{abbrv}

\bibliography{template}

\begin{thebibliography}{10}

\bibitem{amar2005low}
R.~Amar, J.~Eagan, and J.~Stasko.
\newblock Low-level components of analytic activity in information
  visualization.
\newblock In {\em Proceedings of the Proceedings of the 2005 IEEE Symposium on
  Information Visualization}, INFOVIS '05, pages 15--, Washington, DC, USA,
  2005. IEEE Computer Society.

\bibitem{Amar:2004}
R.~Amar and J.~Stasko.
\newblock Best paper: A knowledge task-based framework for design and
  evaluation of information visualizations.
\newblock In {\em IEEE Symposium on Information Visualization}, pages 143--150,
  2004.

\bibitem{bertin:book83}
J.~Bertin.
\newblock {\em Semiology of graphics}.
\newblock University of Wisconsin Press, 1983.

\bibitem{bouali2015vizassist}
F.~Bouali, A.~Guettala, and G.~Venturini.
\newblock Vizassist: an interactive user assistant for visual data mining.
\newblock {\em The Visual Computer}, pages 1--17, 2015.

\bibitem{cleveland1984graphical}
W.~S. Cleveland and R.~McGill.
\newblock Graphical perception: Theory, experimentation, and application to the
  development of graphical methods.
\newblock {\em Journal of the American statistical association},
  79(387):531--554, 1984.

\bibitem{cleveland1985graphical}
W.~S. Cleveland and R.~McGill.
\newblock Graphical perception and graphical methods for analyzing scientific
  data.
\newblock {\em Science}, 229(4716):828--833, 1985.

\bibitem{correll2014error}
M.~Correll and M.~Gleicher.
\newblock Error bars considered harmful: Exploring alternate encodings for mean
  and error.
\newblock {\em IEEE transactions on visualization and computer graphics},
  20(12):2142--2151, 2014.

\bibitem{croxton1927bar}
F.~E. Croxton and R.~E. Stryker.
\newblock Bar charts versus circle diagrams.
\newblock {\em Journal of the American Statistical Association},
  22(160):473--482, 1927.

\bibitem{dambacher2016graphs}
M.~Dambacher, P.~Haffke, D.~Gro{\ss}, and R.~H{\"u}bner.
\newblock Graphs versus numbers: How information format affects risk aversion
  in gambling.
\newblock {\em Judgment and Decision Making}, 11(3):223, 2016.

\bibitem{TableauData}
T.~Datasets.
\newblock https://public.tableau.com/s/resources, 2015.

\bibitem{demiralp2017foresight}
C.~Demiralp, P.~J. Haas, S.~Parthasarathy, and T.~Pedapati.
\newblock Foresight: Recommending visual insights.
\newblock {\em Proc. VLDB Endow.}, 10(12):1937--1940, 2017.

\bibitem{Dimara}
E.~Dimara, A.~Bezerianos, and P.~Dragicevic.
\newblock Conceptual and methodological issues in evaluating multidimensional
  visualizations for decision support.
\newblock {\em IEEE Transactions on Visualization and Computer Graphics},
  24(1):749--759, Jan 2018.

\bibitem{eells1926relative}
W.~C. Eells.
\newblock The relative merits of circles and bars for representing component
  parts.
\newblock {\em Journal of the American Statistical Association},
  21(154):119--132, 1926.

\bibitem{ehrenberg1975data}
A.~E. Ehrenberg.
\newblock {\em Data Reduction: Analysing and interpreting statistical data}.
\newblock John Wiley and Sons, London, 1975.

\bibitem{few2006information}
S.~Few.
\newblock {\em Information dashboard design}.
\newblock O'Reilly, 2006.

\bibitem{garcia2010profits}
R.~Garcia-Retamero and M.~Galesic.
\newblock Who proficts from visual aids: Overcoming challenges in people's
  understanding of risks.
\newblock {\em Social science \& medicine}, 70(7):1019--1025, 2010.

\bibitem{harrison2014ranking}
L.~Harrison, F.~Yang, S.~Franconeri, and R.~Chang.
\newblock Ranking visualizations of correlation using weber's law.
\newblock {\em Visualization and Computer Graphics, IEEE Transactions on},
  20(12):1943--1952, 2014.

\bibitem{henderson1981building}
H.~V. Henderson and P.~F. Velleman.
\newblock Building multiple regression models interactively.
\newblock {\em Biometrics}, 37(2):391--411, 1981.

\bibitem{Kandel:2012}
S.~Kandel, R.~Parikh, A.~Paepcke, J.~M. Hellerstein, and J.~Heer.
\newblock Profiler: Integrated statistical analysis and visualization for data
  quality assessment.
\newblock In {\em Proceedings of the International Working Conference on
  Advanced Visual Interfaces}, AVI '12, pages 547--554, New York, NY, USA,
  2012. ACM.

\bibitem{2016-beyond-webers-law}
M.~Kay and J.~Heer.
\newblock Beyond weber's law: A second look at ranking visualizations of
  correlation.
\newblock {\em IEEE Transactions on Visualization and Computer Graphics},
  22(1):469--478, Jan 2016.

\bibitem{kay2016ish}
M.~Kay, T.~Kola, J.~R. Hullman, and S.~A. Munson.
\newblock When (ish) is my bus? user-centered visualizations of uncertainty in
  everyday, mobile predictive systems.
\newblock In {\em Proceedings of the 2016 CHI Conference on Human Factors in
  Computing Systems}, pages 5092--5103. ACM, 2016.

\bibitem{kosslyn1989understanding}
S.~M. Kosslyn.
\newblock Understanding charts and graphs.
\newblock {\em Applied cognitive psychology}, 3(3):185--225, 1989.

\bibitem{LeeVlat}
S.~Lee, S.~H. Kim, and B.~C. Kwon.
\newblock Vlat: Development of a visualization literacy assessment test.
\newblock {\em IEEE Transactions on Visualization and Computer Graphics},
  PP(99):1--1, 2016.

\bibitem{Mackinlay:2007}
J.~Mackinlay, P.~Hanrahan, and C.~Stolte.
\newblock Show me: Automatic presentation for visual analysis.
\newblock {\em IEEE Transactions on Visualization and Computer Graphics},
  13(6):1137--1144, Nov. 2007.

\bibitem{olson2014ways}
J.~S. Olson and W.~A. Kellogg.
\newblock {\em Ways of Knowing in HCI}.
\newblock Springer, 2014.

\bibitem{Pandey:2016}
A.~V. Pandey, J.~Krause, C.~Felix, J.~Boy, and E.~Bertini.
\newblock Towards understanding human similarity perception in the analysis of
  large sets of scatter plots.
\newblock In {\em Proceedings of the 2016 CHI Conference on Human Factors in
  Computing Systems}, CHI '16, pages 3659--3669, New York, NY, USA, 2016. ACM.

\bibitem{pinker1990theory}
S.~Pinker.
\newblock A theory of graph comprehension.
\newblock {\em Artificial intelligence and the future of testing}, pages
  73--126, 1990.

\bibitem{UCIDataset}
U.~M.~L. Repository.
\newblock https://archive.ics.uci.edu/ml/datasets.html, 2016.

\bibitem{saket2017visualization}
B.~Saket, H.~Kim, E.~T. Brown, and A.~Endert.
\newblock Visualization by demonstration: An interaction paradigm for visual
  data exploration.
\newblock {\em IEEE Transactions on Visualization \& Computer Graphics},
  (1):331--340, 2017.

\bibitem{saket2017evaluating}
B.~Saket, A.~Srinivasan, E.~D. Ragan, and A.~Endert.
\newblock Evaluating interactive graphical encodings for data visualization.
\newblock {\em IEEE Transactions on Visualization and Computer Graphics}, 2017.

\bibitem{santos2008evaluating}
B.~S. Santos.
\newblock Evaluating visualization techniques and tools: What are the main
  issues.
\newblock {\em The AVI Workshop on Beyond Time and Errors: Novel Evaluation
  Methods For information Visualization (BELIV '08)}, 2008.

\bibitem{siegrist1996use}
M.~Siegrist.
\newblock The use or misuse of three-dimensional graphs to represent
  lower-dimensional data.
\newblock {\em Behaviour \& Information Technology}, 15(2):96--100, 1996.

\bibitem{simkin1987information}
D.~Simkin and R.~Hastie.
\newblock An information-processing analysis of graph perception.
\newblock {\em Journal of the American Statistical Association},
  82(398):454--465, 1987.

\bibitem{skau2016arcs}
D.~Skau and R.~Kosara.
\newblock Arcs, angles, or areas: Individual data encodings in pie and donut
  charts.
\newblock In {\em Computer Graphics Forum}, volume~35, pages 121--130. Wiley
  Online Library, 2016.

\bibitem{spence1991displaying}
I.~Spence and S.~Lewandowsky.
\newblock Displaying proportions and percentages.
\newblock {\em Applied Cognitive Psychology}, 5(1):61--77, 1991.

\bibitem{SpotFire}
SpotFire.
\newblock http://www.spotfire.com, 2016.

\bibitem{eurovisshort.20171133}
A.~Srinivasan and J.~T. Stasko.
\newblock {Natural Language Interfaces for Data Analysis with Visualization:
  Considering What Has and Could Be Asked}.
\newblock In B.~Kozlikova, T.~Schreck, and T.~Wischgoll, editors, {\em EuroVis
  2017 - Short Papers}. The Eurographics Association, 2017.

\bibitem{articulate}
Y.~Sun, J.~Leigh, A.~Johnson, and S.~Lee.
\newblock Articulate: A semi-automated model for translating natural language
  queries into meaningful visualizations.
\newblock In R.~Taylor, P.~Boulanger, A.~Kr{\"u}ger, and P.~Olivier, editors,
  {\em Smart Graphics}, pages 184--195. Springer Berlin Heidelberg, 2010.

\bibitem{Tableau}
Tableau.
\newblock Tableau software, http://www.tableau.com/, 2016.

\bibitem{Talbot_bar}
J.~Talbot, V.~Setlur, and A.~Anand.
\newblock Four experiments on the perception of bar charts.
\newblock {\em IEEE Transactions on Visualization and Computer Graphics},
  20(12):2152--2160, Dec 2014.

\bibitem{vartak2014seedb}
M.~Vartak, S.~Madden, A.~Parameswaran, and N.~Polyzotis.
\newblock Seedb: Automatically generating query visualizations.
\newblock {\em Proc. VLDB Endow.}, 7(13):1581--1584, Aug. 2014.

\bibitem{2015-voyager}
K.~Wongsuphasawat, D.~Moritz, A.~Anand, J.~Mackinlay, B.~Howe, and J.~Heer.
\newblock Voyager: Exploratory analysis via faceted browsing of visualization
  recommendations.
\newblock {\em IEEE Transactions on Visualization and Computer Graphics},
  22(1):649--658, Jan 2016.

\bibitem{zacks1999bars}
J.~Zacks and B.~Tversky.
\newblock Bars and lines: A study of graphic communication.
\newblock {\em Memory \& Cognition}, 27(6):1073--1079, 1999.

\end{thebibliography}

%


\begin{IEEEbiographynophoto}{Bahador Saket}
is currently a Ph.D. student at Georgia Institute of Technology. His research focuses on the design of interaction techniques for visual data exploration. He is also interested in conducting experiments as a method to understand how visualizations can be used to support data analysis.
\end{IEEEbiographynophoto}

\begin{IEEEbiographynophoto}{Alex Endert} is an Assistant Professor in the School of Interactive Computing at Georgia Tech. He directs the Visual Analytics Lab, where him and his students explore novel user interaction techniques for visual analytics. His lab often applies these fundamental advances to domains including text analysis, intelligence analysis, cyber security, decision-making, and others. He received his Ph.D. in Computer Science at Virginia Tech in 2012. 
\end{IEEEbiographynophoto}

\begin{IEEEbiographynophoto}{\c{C}a\u{g}atay Demiralp} is a research scientist at IBM and the co-founder and chief scientific advisor at Fitnescity. His current research focuses around two themes: 1) Automating visual data exploration for scalable guided data analysis and 2) improving the data science pipeline with interactive tools that facilitate iterative visual data and model experimentation.
Before IBM, \c{C}a\u{g}atay was a postdoctoral scholar at Stanford University and member of the Interactive Data Lab at the University of Washington. He obtained his PhD from Brown University.

\end{IEEEbiographynophoto}




\end{document}